# Dynamical system-based computational models for solving combinatorial optimization on hypergraphs

Mohammad Khairul Bashar, Antik Mallick, Avik W. Ghosh, Nikhil Shukla

Department of Electrical & Computer Engineering, University of Virginia, Charlottesville,

VA- 22904, USA

*E-mail: ns6pf@virginia.edu

**Abstract:**

The intrinsic energy minimization in dynamical systems offers a valuable tool for minimizing the objective functions of computationally challenging problems in combinatorial optimization. However, most prior works have focused on mapping such dynamics to combinatorial optimization problems whose objective functions have quadratic degree (e.g., MaxCut); such problems can be represented and analyzed using graphs. However, the work on developing such models for problems that need objective functions with degree greater than two, and subsequently, entail the use of hypergraph data structures, is relatively sparse. In this work, we develop dynamical system-inspired computational models for several such problems. Specifically, we define the 'energy function' for hypergraph-based combinatorial problems ranging from Boolean SAT and its variants to integer factorization, and subsequently, define the resulting system dynamics. We also show that the design approach is applicable to optimization problems with quadratic degree, and use it develop a new dynamical system formulation for minimizing the Ising Hamiltonian. Our work not only expands on the scope of problems that can be directly mapped to, and solved using physics-inspired models, but also creates new opportunities to design high-performance accelerators for solving combinatorial optimization.

Despite the tremendous strides achieved across the entire digital hardware-software ecosystem, certain combinatorial optimization problems are still considered challenging to solve using digital computers. Such problems belong to the NP hard computational complexity class. This has motivated the exploration of many alternate computing models and approaches spanning from quantum computing [1]-[3] to classical analog methods using dynamical systems such as neural networks [4]-[6] and oscillator networks [7]-[16]. The analog computing approach, focus of the present work, exploits the fact that combinatorial optimization problems entail the minimization of an objective function, and thus, exhibit a natural similarity to the minimization of energy in a dynamical system. Consequently, this has motivated the formulation of physics-based computational models [7], [17]-[22], inspired by dynamical systems, for solving such problems as well as others [23]-[26]. For instance, Wang *et al.,* [7] showcased how the challenge of minimizing the Ising Hamiltonian (and the equivalent MaxCut problem) can be formulated in terms of the dynamics of coupled oscillators under second harmonic injection. Furthermore, we recently formulated models, inspired by dynamical systems of synchronized oscillators, for solving other combinatorial optimization problems such as the Max-K-Cut, the Traveling Salesman Problem (TSP), the graph partitioning problem among others [27]. Most importantly, all of these problems share a common property that they can be expressed using objective functions that have quadratic degree [28].

However, there is a larger class of problems such as Boolean satisfiability and integer factorization among others wherein the objective functions have a degree greater than *two*. Such problems entail the use of hypergraphs for their representation and analysis. A hypergraph can be considered as a generalization of graphical data structures wherein an edge (known as a hyperedge) can connect any number of vertices; this is in contrast to a graph where an edge can join a maximum of two vertices. Analog models for solving combinatorial problems in hypergraphs have been relatively less explored [29]-[31]. We note that such problems can, in theory, be reduced to problems that have objective functions with quadratic degree [32], [33]. However, this typically involves the introduction of additional auxiliary variables which can effectively increase the size of the (quadratic degree) combinatorial problem that must then be solved [34], [35]. Therefore, our goal in this work is to formulate analog computational models for solving such problems without introducing auxiliary variables. We would also like to clarify that the emphasis of the current work lies on formulating the computational models, and not necessarily on the physical implementation of the dynamical system.

Our approach builds on the foundational work performed by Ercsey-Ravasz *et al.* [29], wherein the authors proposed an approach for solving the Boolean Satisfiability (SAT) problem using continuous (analog) variables. The SAT problem is defined as the challenge of evaluating a Boolean assignment (1 or 0) that will satisfy a Boolean formula expressed in the conjunctive normal form (CNF); $Y = C_1 \wedge C_2 \wedge ... C_M$. The decision version of the problem evaluates if such an assignment exists. Building on this elegant method, here, we formulate computational models for: (a) the NAE (Not-All-Equal) SAT problem, which is an NP-complete variant of the SAT problem. Besides finding an assignment for the Boolean variables such that every clause is satisfied, the NAE-SAT problem also requires that at least one literal in every clause is false. Further, the computational model for the NAE-SAT problem can be extended to the Set Splitting problem, which evaluates if there is exists a partition that splits a finite set into two parts such that all the subsets of the finite set are split by the partition. The Set Splitting problem is a special case of the NAE-SAT problem wherein all the variables in the normal form (positive NAE-SAT); (b) Integer factorization problem, considered here as the problem of dividing a number into two integer factors; (c) The Graph Isomorphism problem, which evaluates if two graphs with the same number of edges and vertices (non-trivial case) have the same edge connectivity. (d) Finally, we show that the proposed approach can be used to minimize the Ising Hamiltonian (quadratic problem), and in fact, provides an alternate dynamical system formulation to the well-known oscillator-based dynamical system proposed earlier [7]. Subsequently, using this formulation, we show its application in solving the archetypal Maximum Cut (MaxCut) problem, defined as the challenge of dividing the nodes of a graph into two sets such that the number of shared edges (among the two sets) is maximized.

**SAT**

We first consider the Boolean SAT problem where we represent each variable $x_i$ in the Boolean expression by $\gamma_i \equiv \frac{1+cos(\alpha_i)}{2}$, where $\alpha_i$ is an analog variable. The $cos(.)$ function sets the bounds of $\gamma_i$ to [0,1], and ensures that the Boolean variable and its analog counterpart have the same value at the maxima and the minima of the analog variable function. We note that while the above formulation resembles a (level-shifted) oscillator, the two are not exactly equivalent (the 'ωt' term is not considered here); nevertheless, we will refer to $\alpha$ as a 'phase' for simplicity. For each clause $C_m$, we define $K_m(\alpha) = \prod_{i=1}^{N}\left(1 - \left(\frac{1+c_{mi}\cos(\alpha_i)}{2}\right)\right)$. $c_{mi} = 1(-1)$, if $x_i$ in the $m^{th}$ clause appears in the normal (negated) form, respectively; $c_{mi} = 0$, if $x_i$ is absent from the $m^{th}$ clause. $K_m(\alpha)$ can be considered as an analog equivalent of $1 - C_m$, and exhibits the property that $K_m = 0$ if and only if the clause is satisfied ($C_m = 1$) i.e., at least one variable is TRUE. We define a continuous time dynamical system given by $\frac{d\alpha}{dt} = F = (-\nabla_\alpha V)$, which has an energy function given by:

$$V = \sum_{m=1}^{M} A(K_m(\alpha))^2 \tag{1}$$

where $A$ ($> 0$) is a constant. It can be observed from equation (1) that $V$ is minimized by maximizing the number of satisfied clauses. Further, $V = 0$ is the global minima of the function and is attained when all the clauses are satisfied i.e., $K_m = 0$ for m =1, 2,…, M.

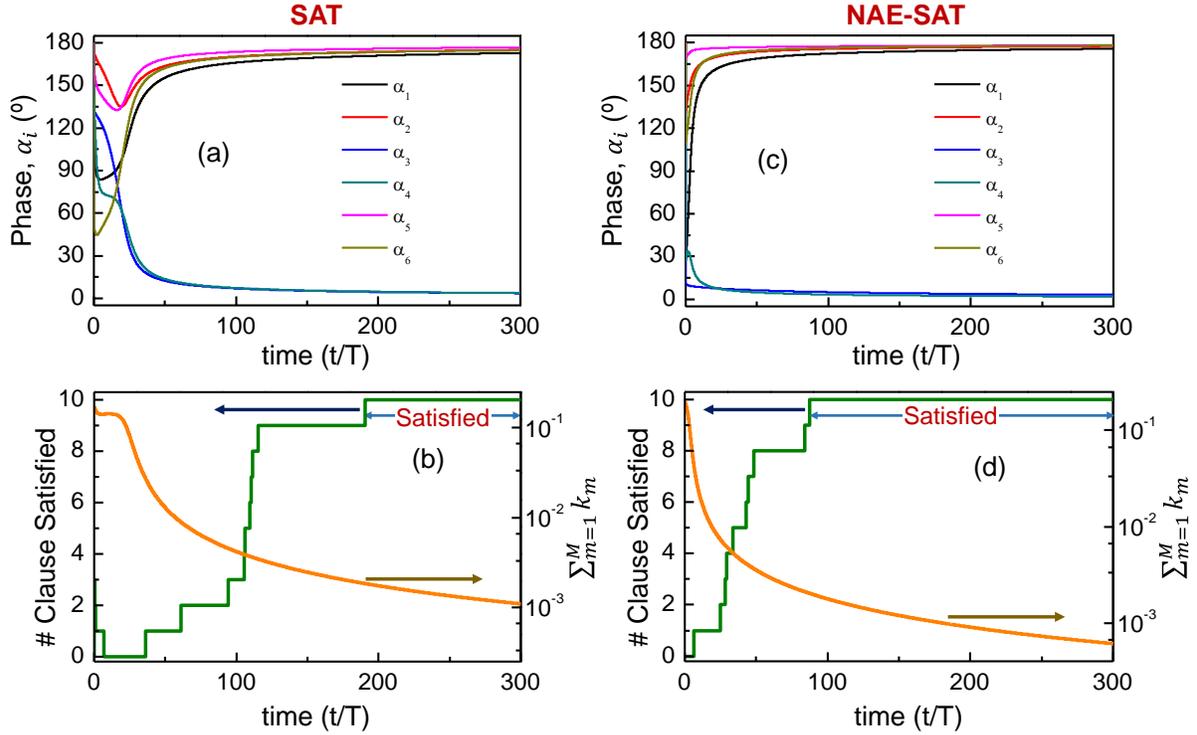

**Fig. 1.** Evolution of the phases ($\alpha_i$), $K_m$ ($K_{m,NAE}$) & the number of satisfied clauses, respectively, as a function of time for: (a)(b) the Boolean SAT problem; and (c)(d) the Boolean NAE-SAT problem. The Boolean expression considered in this illustrative example is given in Appendix II. Since numerical methods are used for solving the dynamics, a threshold value of $K_m(K_{m,NAE}) < 4 \times 10^{-4}$ was used for a clause to be considered as TRUE (satisfied).

Appendix I shows that $\frac{dV}{dt} \leq 0$ implying that the system always evolves to minimize $V$ (energy), or in other words, maximize the number satisfied clauses.

The corresponding dynamics can be computed as:

$$\frac{d\alpha_i}{dt} = (-\nabla_\alpha V)_i = \sin(\alpha_i) \cdot \left(-\sum_{m=1}^{M}\left[2AK_m(\alpha)\left[\frac{c_{mi}K_m(\alpha)}{1 - c_{mi}\cos(\alpha_i)}\right]\right]\right) \quad (2)$$

Fig. 1(a),(b) show a representative Boolean SAT problem solved using the above computational model. It can be observed that the system minimizes $K_m(\alpha)$ which subsequently maximizes the number of clauses satisfied. We once again acknowledge that this formulation derives strong inspiration from the elegant analog dynamics formulated by Ercsey-Ravasz *et al.* [29], and serves as the building block for the dynamical system-based computational models developed for other problems in this work.

**NAE-SAT:**

The NAE-SAT problem is an NP-complete variant of the SAT problem with the added constraint that every clause must contain a literal that is true and false. To evaluate the NAE-SAT problem for a Boolean expression $Y = C_1 \wedge C_2 \wedge ... C_M$, each clause $C_i = (x_1 \vee \overline{x_2} \vee x_3 \vee ... \vee x_N)$ in the original expression can be modified to $C_{NAE,i} = (x_1 \vee \overline{x_2} \vee x_3 \vee ... \vee x_N) \cdot \overline{(x_1 \wedge \overline{x_2} \wedge x_3 \wedge ... \wedge x_N)} \equiv C_i \cdot S_i$, where $S_i$ is the negation of the conjunction of all the literals in that clause. While $C_i$ imposes the condition that at least one literal must be true, $S_i$ imposes the added constraint that at least one literal must be false in order that $C_{NAE,i} = 1$ (TRUE); an example for this is shown in Appendix III. Thus, the NAE-SAT problem can be expressed as evaluating if the expression $Y_{NAE} = C_{NAE,1} \wedge C_{NAE,2} \wedge$

... $C_{NAE,M}$ can be made TRUE. To define the computational model for this problem, we again define an energy function similar to that of the SAT problem:

$$V = \sum_{m=1}^{M} A \left(K_{m,NAE}(\alpha)\right)^2 \tag{3}$$

albeit with a different analog formulation for each clause. $K_{m,NAE}(\alpha)$ is now defined as:

$$K_{m,NAE}(\alpha) = \left[\prod_{i=1}^{N}\left(1 - \left(\frac{1 + c_{mi}\cos(\alpha_i)}{2}\right)\right)\right] + \left[\prod_{i=1}^{N}\left(\frac{1 + c_{mi}\cos(\alpha_i)}{2}\right)\right] \tag{4a}$$

$$K_{m,NAE}(\alpha) = K_m^1(\alpha) + K_m^2(\alpha) \tag{4b}$$

Here, $K_m^1(\alpha)$ is similar to the $K_m(\alpha)$ defined for the SAT problem, and essentially captures the constraint that the contribution of that clause to the energy function is zero when the clause is satisfied. $K_m^2(\alpha)$ is formulated to define the additional constraint for the NAE-SAT problem entailing that all the literals cannot be equal to each other. Together, the formulation of $K_{m,NAE}(\alpha)$ for the NAE SAT clause ensures that it's contribution to the energy function is zero only when the clause is satisfied i.e., at least one literal is true, and all the literals are not equal to each other. The latter condition essentially ensures that at least one literal must be false.

The corresponding system dynamics can be defined by:

$$\frac{d\alpha_i}{dt} = \left(-\nabla_\alpha V(\alpha)\right)_i = -\sum_{m=1}^{M} 2A\, K_{m,NAE}(\alpha)\left(\frac{dK_{m,NAE}(\alpha)}{d\alpha_i}\right) \tag{5a}$$

where,

$$\frac{dK_m(\alpha_i)}{d\alpha_i} = \frac{-c_{mi}K_m^1(\alpha)}{1-c_{mi}\cos(\alpha_i)} \cdot (-\sin(\alpha_i)) + \frac{c_{mi}K_m^2(\alpha)}{1+c_{mi}\cos(\alpha_i)} \cdot (-\sin(\alpha_i)) \quad (5b)$$

$$= \left[\frac{-c_{mi}K_m^1(\alpha)}{1-c_{mi}\cos(\alpha_i)} + \frac{c_{mi}K_m^2(\alpha)}{1+c_{mi}\cos(\alpha_i)}\right] \cdot (-\sin(\alpha_i)) \quad (5c)$$

$$\frac{d\alpha_i}{dt} = \sin(\alpha_i)\left(-\sum_{m=1}^{M} 2A\,K_m(\alpha)\left[\frac{c_{mi}K_m^1(\alpha)}{1-c_{mi}\cos(t+\alpha_i)} - \frac{c_{mi}K_m^2(\alpha)}{1+c_{mi}\cos(t+\alpha_i)}\right]\right) \quad (5d)$$

Fig. 1(c),(d) show a representative example of a NAE-SAT expression solved using the above computing model.

**Set Splitting:**

Given a finite set S where $S_1$, $S_2$… $S_N$ are the subsets, the objective of the Set Splitting problem is to evaluate if there exists a partition that divides all the subsets into two parts. This problem is equivalent to computing the solution of the positive NAE-SAT i.e., with only normal variables. To establish the relationship between the Set Splitting problem and the NAE-SAT problem, each element in the set can be represented by a variable $x_i$; $x_i = 1(0)$, if $x_i$ lies in Set I (II) (or vice-versa). We note that only variables in the normal form are needed. Subsequently, each subset $S_i$ of the finite set can be mapped to $C_{NAE,i}$. It can be observed that only if the set is split (i.e., some nodes of $S_i$ lie in Set I and II each) by the partition, $C_{NAE,i}$ evaluates to 1; if the nodes of a subset $S_i$ lie entirely in Set I or II, $C_{NAE,i} = 0$. A partition that splits all the subsets exists when all $C_{NAE}$ are satisfied i.e., $V = 0$.

**Integer Factorization:**

The integer factorization problem is an NP complete problem that entails finding the integer factors of a number. Here, we consider the challenge of dividing a number $Z$ into two factors $X$ and $Y$ such that $XY = Z$, or in other words, $XY - Z = 0$. Expressing the factors $X$ and $Y$ in binary form, this relationship can be used to formulate an energy function:

$$V = A\left(\left(\sum_{i=1}^{N} 2^{i-1}\left(\frac{1+\tanh(k\cos(\alpha_i))}{2}\right)\right)\left(\sum_{j=N+1}^{2N} 2^{j-N-1}\left(\frac{1+\tanh(k\cos(\alpha_j))}{2}\right)\right) - F\right)^2 \quad (6)$$

where each binary bit in $X$ and $Y$ is represented by $\left(\frac{1+\tanh(k\cos(\alpha_{i,j}))}{2}\right)$; here, $\alpha_i$ and $\alpha_j$ are used to represent the bits in $X$ and $Y$, respectively, and $k$ essentially decides the 'steepness' of the $\tanh(.)$ function. This formulation of the energy function is inspired from that adopted by Borders *et al.* [36] and it can be observed that the energy function is expressed as a 'product of sums', instead of the 'sum of products' used in the formulation for the SAT and the NAE-SAT problems. The corresponding system dynamics are given by:

$$\frac{d\alpha_i}{dt} = \left(-\nabla_\alpha V(\alpha)\right)_i = \quad (7a)$$

$$\left( \sin(\alpha_i) \left( 2A \left( \left( \sum_{j=1}^{N} 2^{j-1} \left( \frac{1 + \tanh(k\cos(\alpha_j))}{2} \right) \right) \left( \sum_{m=N+1}^{2N} 2^{m-N-1} \left( \frac{1 + \tanh(k\cos(\alpha_m))}{2} \right) \right) \right. \right. \right.$$

$$\left. \left. \left. - F \right) \left( \sum_{n=N+1}^{2N} 2^{n-N-1} \left( \frac{1 + \tanh(k\cos(\alpha_n))}{2} \right) \right) \cdot 2^{i-2} \cdot k \cdot \text{sech}^2(k\cos(\alpha_i)) \right) \right)$$

$$\frac{d\alpha_j}{dt} = \left( -\nabla_\alpha V(\alpha) \right)_j = \tag{7b}$$

$$\left( \sin(\alpha_j) \left( 2A \left( \left( \sum_{i=1}^{N} 2^{i-1} \left( \frac{1 + \tanh(k\cos(\alpha_i))}{2} \right) \right) \left( \sum_{m=N+1}^{2N} 2^{m-N-1} \left( \frac{1 + \tanh(k\cos(\alpha_m))}{2} \right) \right) \right. \right. \right.$$

$$\left. \left. \left. - F \right) \left( \sum_{n=1}^{N} 2^{n-1} \left( \frac{1 + \tanh(k\cos(\alpha_n))}{2} \right) \right) \cdot 2^{j-N-2} \cdot k \cdot \text{sech}^2(k\cos(\alpha_j)) \right) \right)$$

Fig. 2 presents an illustrative example showing the integer factorization of 899 performed using the above model. We note that the $\tanh(.)$ function used in the analog formulation of the bits of the factors $X$ and $Y$ helps to effectively 'binarize' the output of the $\cos(.)$ function. This is because the energy function (without the $\tanh(.)$ function) may not always converge to *integer* factors of $Z$ i.e., $V = 0$, may also be achieved when $\cos(\alpha_{i,j}) \neq 1$ or $-1$, resulting in non-integer factors. The $\tanh(.)$ function helps drive the phases towards 0 ($\cos(\alpha_{i,j}) = 1$) or π ($\cos(\alpha_{i,j}) = -1$). This can be understood by considering the $\text{sech}^2(.)$ function (arising from the $\tanh(.)$ term in the energy function) in the resulting dynamical system (equation (7b)) – the $\text{sech}^2(.)$ function achieves a maximum (=1) when the (resulting) input to the function is zero (i.e., $\cos(\alpha_{i,j}) = 0$; $\alpha_{i,j} = \pm\frac{\pi}{2}$, and the corresponding 'bit' achieves a value of 0.5), and decays asymptotically towards zero as the input deviates from zero (i.e., $sech^2(.)$ reduces as $\alpha_{i,j} \to 0$ $(cos(\alpha_{i,j}) \to 1)$ and $\alpha_{i,j} \to \pi$ $(cos(\alpha_{i,j}) \to -1)$. This implies that the function selectively reduces the perturbation as phases settle towards $\alpha_{i,j} = 0$ and π. This impact of using the $tanh(.)$ function is illustrated with an example in Appendix IV.

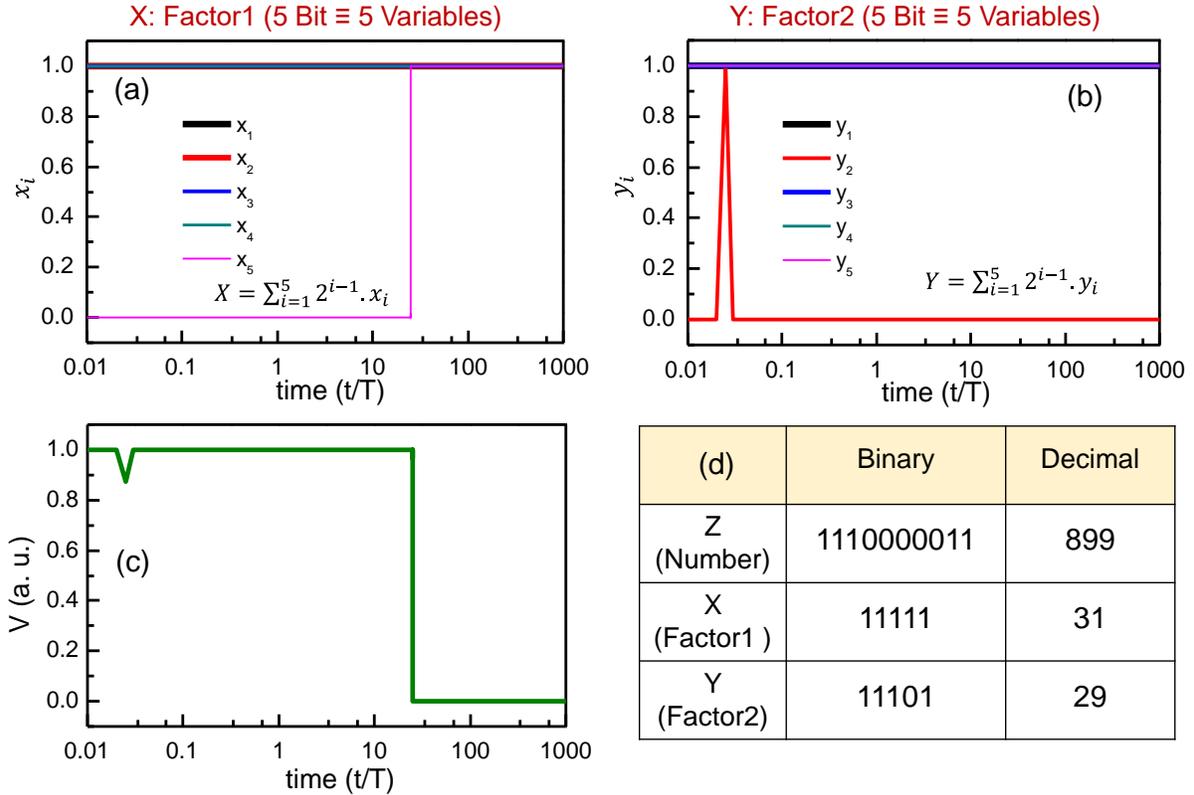

**Fig. 2.** Integer factorization of 899. Temporal evolution of: (a) (b) the variables corresponding to bits in the integer factors X and Y, respectively; (c) Energy (V). (d) Integer factors X and Y computed by the system expressed in binary and decimal form.

**Graph Isomorphism:**

This problem is defined as the challenge of evaluating if two graphs are equivalent. The non-trivial case entails evaluating if two graphs with equal number of vertices and edges have the same edge connectivity i.e., adjacency matrices. Mathematically this problem can be expressed as: Given two graphs with adjacency matrices defined by $A$ and $B$, is there a permutation matrix $P$ such that $AP = PB$? [37] To formulate the computational model for this problem, we represent each element in $P$ as $P_{ij} = \left(\frac{1+tanh(k\,cos(\alpha_{ij}))}{2}\right)$, and formulate the energy function as:

$$V = \sum_{m=1}^{N}\sum_{n=1}^{N} A(K_{mn}(\alpha))^2 \quad (8)$$

Here, $N \times N$ is the size of the matrices $A$ and $B$. $K_{mn}$ is defined as:

$$K_{mn} = \frac{1}{N}\left(\sum_{r=1}^{N} a_{mr}\left(\frac{1+tanh(k\,cos(\alpha_{rn}))}{2}\right) - \sum_{s=1}^{N}\left(\frac{1+tanh(k\,cos(\alpha_{ms}))}{2}\right)b_{sn}\right) \quad (9)$$

and represents the element-wise difference between the products of $AP$ and $PB$ i.e., $AP - PB$; see Appendix V for details on the derivation of $K_{mn}$. $K_{mn} = 0$ when the two terms are equal, and $V = 0$ when all the terms (element wise) are matched. We note here that the energy function has quadratic degree. Nevertheless, the problem is considered since the formulation is well aligned to the dynamical system proposed here. The corresponding dynamics of the system can be defined by:

$$\frac{d\alpha_{ij}}{dt} = (-\nabla_\alpha V(\alpha))_{ij} = -\sum_{m=1}^{N}\sum_{n=1}^{N} 2AK_{mn}(\alpha)\left(\frac{dK_{mn}(\alpha)}{d\alpha_{ij}}\right) \quad (10a)$$

where,

$$\frac{dK_{mn}(\alpha)}{d\alpha_{ij}} = -\frac{1}{2N}\sin(\alpha_{ij}).k\,sech^2\left(kcos(\alpha_{ij})\right).\left[(a_{mi})_{n=j} - (b_{jn})_{m=i}\right] \quad (10b)$$

$$\frac{d\alpha_{ij}}{dt} = sin(\alpha_{ij})\left[\frac{A}{N} k\, sech^2\left(k\, cos(\alpha_{ij})\right) \cdot \left(\sum_{m=1}^{N} a_{mi}k_{mj} - \sum_{n=1}^{N} b_{jn}k_{in}\right)\right] \qquad (10c)$$

Fig. 3 shows an illustrative example evaluating the isomorphism between two graphs using the model proposed above.

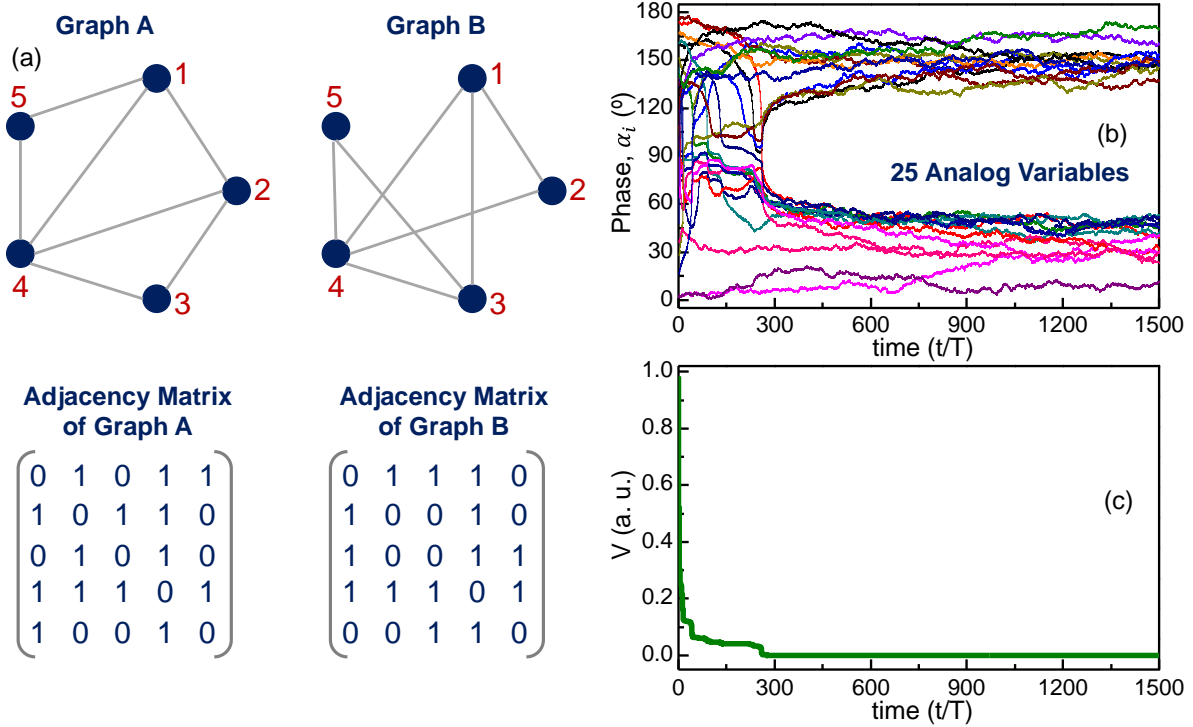

**Fig. 3.** (a) Two representative graphs along with their respective adjacency matrices; (b)(c) Evolution of the phases and the total energy as a function of time, respectively. It can be observed that the energy (V) reduces to 0 indicating that the graphs are isomorphic.

**Minimization of the Ising Hamiltonian and MaxCut**

Next, we also illustrate how the above approach can be applied to minimizing the Ising Hamiltonian, and subsequently, show its application in solving the Maximum Cut problem- the minima of the Ising Hamiltonian $-\sum_{i,j;i<j}^{N} J_{ij}\sigma_i\sigma_j$ (Zeeman term neglected here) corresponds to the MaxCut of the equivalent graph when an edge between the nodes $i$ and $j$ is represented by $J_{ij} = -1$. Thus, both the problems also have objective functions with quadratic degree. We formulate the energy function for the above problem as:

$$V = A \sum_{i,j;i \neq j}^{N} J_{ij}(\cos(\alpha_i) - \cos(\alpha_j))^2 \tag{11}$$

where $J_{ij} = -1(0)$, if an edge is present (absent) between the nodes $i$ and $j$, respectively. The energy function in equation (11) can be expressed as:

$$V = A \sum_{i,j;i \neq j}^{N} J_{ij}(\cos(\alpha_i))^2 + A \sum_{i,j;i \neq j}^{N} J_{ij}(\cos(\alpha_j))^2 - 2A \sum_{i,j;i \neq j}^{N} J_{ij}\cos(\alpha_i)\cos(\alpha_j) \tag{12a}$$

Further,

$$\sum_{i,j;i \neq j}^{N} J_{ij}(\cos(\alpha_i))^2 = -\sum_{i=1}^{N} \Delta_i(\cos(\alpha_i))^2 \tag{12b}$$

Where $\Delta_i$ is the degree of the $i^{th}$ node in the graph. Therefore, equation (12a) can be expressed as:

$$V = -2A \sum_{i=1}^{N} \Delta_i(\cos(\alpha_i))^2 - 2A \sum_{i,j;i \neq j}^{N} J_{ij}\cos(\alpha_i)\cos(\alpha_j) \tag{12c}$$

Generalizing equation (12c),

$$V = -\sum_{i=1}^{N} C_i(\cos(\alpha_i))^2 - C \sum_{i,j;i \neq j}^{N} J_{ij}\cos(\alpha_i)\cos(\alpha_j) \tag{12d}$$

Where $C_i$ and $C$ are positive constants. It can be observed from equation (12d) that $V$ attains a minimum when $(\alpha_i, \alpha_j) = (0, \pi) \; or \; (\pi, 0)$. At these specific phase points, equation (12d) can be simplified as:

$$V = -\sum_{i=1}^{N} C_i - C \sum_{i,j;i<j}^{N} J_{ij}\cos(\alpha_i)\cos(\alpha_j) \tag{12e}$$

The first term on the right-hand side is essentially a constant for a given graph. Further, by considering each oscillator $\cos(\alpha_i)$ as a spin $\sigma_i$, equation (12e) can be recast as:

$$V = -C \sum_{i,j;i<j}^{N} J_{ij}\sigma_i\sigma_j - C_s \tag{12f}$$

where, $C$ and $C_s$ are positive constants. Equation (12f) is equivalent to the Ising Hamiltonian (the Zeeman term has been neglected here) with a constant offset.

Using equation (12d), the corresponding system dynamics can be defined as:

$$\frac{d\alpha_i}{dt} = \left(-\nabla_\alpha V(\alpha)\right)_i = -2C_i\cos(\alpha_i)\sin(\alpha_i) - C \sum_{j=1;j\neq i}^{N} J_{ij}\sin(\alpha_i)\cos(\alpha_j) \tag{13a}$$

Exploiting the trigonometric relationships: $2\cos(\alpha_i)\sin(\alpha_i) = \sin(2\alpha_i)$, and $2\sin(\alpha_i)\cos(\alpha_j) = \sin(\alpha_i + \alpha_j) + \sin(\alpha_i - \alpha_j)$, equation (13a) can be expressed as:

$$\frac{d\alpha_i}{dt} = -C_i\sin(2\alpha_i) - Q \sum_{j=1;j\neq i}^{N} J_{ij}\left(\sin(\alpha_i + \alpha_j) + \sin(\alpha_i - \alpha_j)\right) \tag{13b}$$

where $Q = \frac{C}{2}$.

Equation (13b) reveals the temporal dynamics of the system. In fact, as a computational model, equation (13) presents an alternative dynamical system to the oscillator-based dynamical system formulation proposed earlier [7] (see Appendix VI) - the ground state energy is still equivalent to the global minima of Ising Hamiltonian for both the systems, but they will evolve with a different set of dynamics. Fig. 4 shows the MaxCut computed on an illustrative graph using the proposed approach compared with the oscillator-based model developed earlier. Optimal solutions are observed in both the cases.

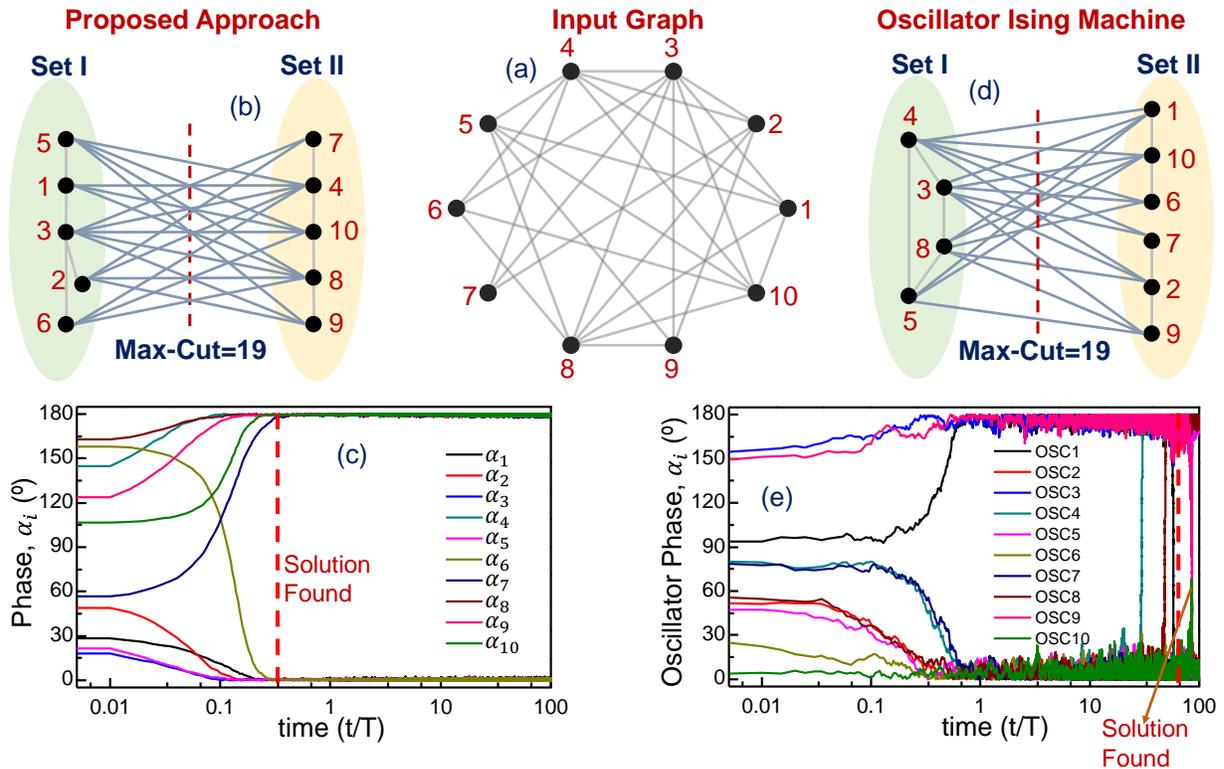

**Fig. 4.** Computing MaxCut. (a) Illustrative graph considered. Phase evolution, and the resulting MaxCut solution computed using the proposed model (b) (c), respectively, and the model developed in [7] ((d) (e)). Optimal solutions are achieved using both the models.

**Conclusion**

This work explores the formulation of computational models inspired by the natural energy minimization in dynamical systems to minimize the objective functions of combinatorial problems with degree greater than two (non-quadratic). In effect, this work helps expands on the scope of problems for which such dynamical system inspired models can be applied. We would like to point out that for problems such as SAT and NAE-SAT, we have considered the decision version of the problems. The above method is applicable to the optimization version (MaxSAT, Max-NAE-SAT) of such problems in the sense that the system will continue to minimize energy which in the case of the SAT and the NAE-SAT problem corresponds to maximizing the number of satisfied clauses. However, since the optimal solution to the problem may not correspond to the global minima of the energy function, the system will be unable to attain steady state. Thus, identifying when the system has converged to the solution is likely to be difficult though there have been recent works on modifying the energy functions to address this challenge [38].

Further, the presence of local minima in the energy function can have a significant impact on the ability of the dynamical system to converge to the optimal solution; the system can get trapped in such local minima resulting in sub-optimal solutions. Here, we would like to point the ground-breaking work by Ercsey-Ravasz *et al.*[29] wherein the authors propose the use of auxiliary variables to ensure convergence to optimal solutions. The impact of incorporating such auxiliary variables on the system dynamics for the various problems considered here will be undertaken in future work. In conclusion, this work expands the applicability of dynamical models to hypergraphs, and bolsters the case for exploring domain specific accelerators that can accelerate such dynamical models for solving graph and hypergraph problems.

**Appendix I**

To show that the energy function defined in equation (1) strictly decreases with time i.e., $\frac{dV}{dt} \leq 0$.

$$\frac{dV}{dt} = \left(2A \sum_{m=1}^{M} K_m(\alpha) \cdot \left[\frac{c_{mi} K_m(\alpha)}{1 - c_{mi} \cos(\alpha_i)}\right] \sin(\alpha_i)\right) \cdot \left(\frac{d\alpha_i}{dt}\right) \quad \text{(A1.1)}$$

Further,

$$\frac{dV}{d\alpha_i} = 2A \sum_{m=1}^{M} K_m(\alpha) \left[\frac{c_{mi} K_m(\alpha)}{1 - c_{mi} \cos(\alpha_i)}\right] \sin(\alpha_i) \quad \text{(A1.2)}$$

It can be observed from equation (A1.1) and (A1.2) that the term in the first bracket on the righthand side of equation (A1.1) is equal to $\frac{dV}{d\alpha_i}$. Further, $\frac{dV}{d\alpha_i} = -\frac{d\alpha_i}{dt}$. Substituting these terms into equation (A1.1), $\frac{dV}{dt}$ can be expressed as

$$\frac{dV}{dt} = -\left(\frac{d\alpha_i}{dt}\right)^2 \leq 0 \quad \text{(A1.3)}$$

Equation (A1.3) implies that the system will always evolve to minimize energy ($V$).

**Appendix II**

The Boolean expression considered for illustrative SAT and NAE-SAT expression is (shown in equation (A2.1), and ass well in Fig. 5 in CNF format):

$$Y = (\overline{x_1} \vee x_2 \vee x_4) \wedge (x_1 \vee x_4 \vee x_5) \wedge (x_2 \vee x_3 \vee x_6) \wedge (\overline{x_2} \vee x_5) \wedge (\overline{x_2} \vee x_6) \wedge (x_1 \vee \overline{x_6}) \wedge (x_2 \vee \overline{x_5} \vee x_6) \wedge (\overline{x_1} \vee \overline{x_4}) \wedge (\overline{x_4} \vee \overline{x_6}) \wedge (\overline{x_5} \vee x_6) \quad (A2.1)$$

| Clause | .CNF Format | | | |
|---|---|---|---|---|
| $C_1$ | -1 | 2 | 4 | 0 |
| $C_2$ | 1 | 4 | 5 | 0 |
| $C_3$ | 2 | 3 | 6 | 0 |
| $C_4$ | -2 | 5 | 0 | |
| $C_5$ | -2 | 6 | 0 | |
| $C_6$ | 1 | -6 | 0 | |
| $C_7$ | 2 | -5 | 6 | 0 |
| $C_8$ | -1 | -4 | 0 | |
| $C_9$ | -4 | -6 | 0 | |
| $C_{10}$ | -5 | 6 | 0 | |

**Fig. 5.** The Boolean expression considered in Fig. 1 expressed in the CNF format.

**Appendix III**

Here, we illustrate with an example, the reformulation of a given clause in a Boolean expression for solving the NAE-SAT problem. Consider a clause $C_i = (x_1 \vee x_2 \vee \overline{x_3} \vee \overline{x_4})$. For the NAE-SAT problem, $C_{NAE-SAT} = C_i S_i$, where $S_i = \overline{(x_1 \wedge x_2 \wedge \overline{x_3} \wedge \overline{x_4})}$. It can be observed from Table I, that the NAE-SAT imposes an additional constraint (highlighted in grey) wherein all the literals cannot be true.

| $x_1$ | $x_2$ | $x_3$ | $x_4$ | $C_i$ (SAT) | $S_i$ | $C_{NAE-SAT} = C_i \cdot S_i$ (NAE-SAT) |
|---|---|---|---|---|---|---|
| 0 | 0 | 0 | 0 | 1 | 1 | 1 |
| 0 | 0 | 0 | 1 | 1 | 1 | 1 |
| 0 | 0 | 1 | 0 | 1 | 1 | 1 |
| 0 | 0 | 1 | 1 | 0 | 1 | 0 |
| 0 | 1 | 0 | 0 | 1 | 1 | 1 |
| 0 | 1 | 0 | 1 | 1 | 1 | 1 |
| 0 | 1 | 1 | 0 | 1 | 1 | 1 |
| 0 | 1 | 1 | 1 | 1 | 1 | 1 |
| 1 | 0 | 0 | 0 | 1 | 1 | 1 |
| 1 | 0 | 0 | 1 | 1 | 1 | 1 |
| 1 | 0 | 1 | 0 | 1 | 1 | 1 |
| 1 | 0 | 1 | 1 | 1 | 1 | 1 |
| **1** | **1** | **0** | **0** | **1** | **0** | **0 (Additional constraint imposed by NAE-SAT)** |
| 1 | 1 | 0 | 1 | 1 | 1 | 1 |
| 1 | 1 | 1 | 0 | 1 | 1 | 1 |
| 1 | 1 | 1 | 1 | 1 | 1 | 1 |

**Table I.** Illustrative example showing reformulation of the clause for NAE-SAT

**Appendix IV**

Here, we evaluate the system dynamics for computing the integer factors of 899 without considering the tanh(.) function in the description of the system energy (equation (6)). The resulting dynamics (without the tanh(.) function) are:

$$\frac{d\alpha_i}{dt} = (-\nabla_\alpha V(\alpha))_i =$$

$$\sin(\alpha_i) \left( 2A \left( \left( \sum_{j=1}^{N} 2^{j-1} \left( \frac{1+\cos(\alpha_j)}{2} \right) \right) \left( \sum_{m=N+1}^{2N} 2^{m-N-1} \left( \frac{1+\cos(\alpha_m)}{2} \right) \right) \right. \right.$$

$$\left. \left. - F \right) \left( \sum_{n=N+1}^{2N} 2^{n-N-1} \left( \frac{1+\cos(\alpha_n)}{2} \right) \right) \cdot 2^{i-2} \right) \qquad (A4.1)$$

$$\frac{d\alpha_j}{dt} = (-\nabla_\alpha V(\alpha))_j =$$

$$\sin(\alpha_j) \left( 2A \left( \left( \sum_{i=1}^{N} 2^{i-1} \left( \frac{1+\cos(\alpha_i)}{2} \right) \right) \left( \sum_{m=N+1}^{2N} 2^{m-N-1} \left( \frac{1+\cos(\alpha_m)}{2} \right) \right) \right. \right.$$

$$\left. \left. - F \right) \left( \sum_{n=1}^{N} 2^{n-1} \left( \frac{1+\cos(\alpha_n)}{2} \right) \right) \cdot 2^{j-N-2} \right) \qquad (A4.2)$$

Fig. 6 shows the resulting dynamics of the system when computing the integer factors of the number 899. It can be observed that without the $tanh(.)$ function, the system can get stabilized when $cos(\alpha) \neq \pm 1$, resulting in non-integer solutions.

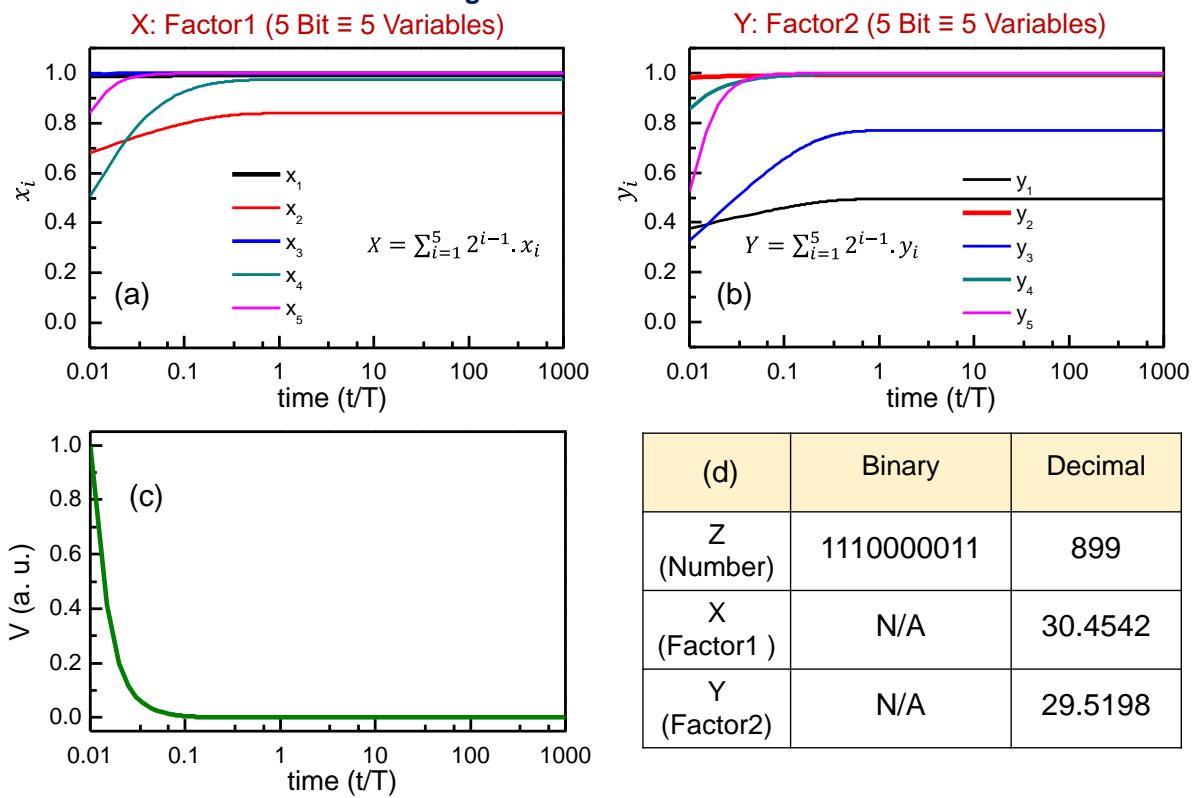

**Fig. 6.** Integer factorization of 899 without using $tanh(.)$ in the energy formulation. Temporal evolution of: (a) (b) the variables representing bits in the integer factors X and Y, respectively; (c) Energy V. (d) Factors X and Y computed by the system. It can be observed that the system settles to non-integer values.

**Appendix V**

Here, we derive the expression for $K_{mn}$ defined in equation (9).

$$K_{mn} = \frac{1}{N}([A][P] - [P][B]) \tag{A5.1}$$

The $ij^{th}$ elements in $[A]$, $[B]$ and $[P]$ are addressed as $a_{ij}, b_{ij}$ and $p_{ij} = \left(\frac{1+tanh(k\,cos(\alpha_{ij}))}{2}\right)$, respectively. Thus, the $mn^{th}$ element in $[X] = [A][P]$ and $[Y] = [P][B]$ can be calculated as:

$$x_{mn} = \sum_{r=1}^{N} a_{mr} \cdot p_{rn} = \sum_{r=1}^{N} a_{mr} \left(\frac{1 + tanh(k\,cos(\alpha_{rn}))}{2}\right) \tag{A5.2}$$

and

$$y_{mn} = \sum_{s=1}^{N} p_{ms} \cdot b_{sn} = \sum_{s=1}^{N} \left(\frac{1 + tanh(k\,cos(\alpha_{ms}))}{2}\right) b_{sn} \tag{A5.3}$$

Subsequently, the $mn^{th}$ element of $[X] - [Y]$ can be computed as,

$$([X] - [Y])_{mn} = \sum_{r=1}^{N} a_{mr} \left(\frac{1 + tanh(k\,cos(\alpha_{rn}))}{2}\right) \tag{A5.4}$$

$$- \sum_{s=1}^{N} \left(\frac{1 + tanh(k\,cos(\alpha_{ms}))}{2}\right) b_{sn}$$

$([X] - [Y])_{mn}$ is normalized to $N$ to calculate $K_{mn}$ used in the energy function,

$$K_{mn} = \frac{1}{N}\left(\sum_{r=1}^{N} a_{mr} \left(\frac{1 + tanh(k\,cos(\alpha_{rn}))}{2}\right) - \sum_{s=1}^{N} \left(\frac{1 + tanh(k\,cos(\alpha_{ms}))}{2}\right) b_{sn}\right) \tag{A3.5}$$

**Appendix VI**

The dynamical system for the oscillator based Ising machine [7] is given by:

$$\frac{d\phi_i(t)}{dt} = -C_1 \sum_{j=1,\ j\neq i}^{N} J_{ij} \sin(\Delta\phi_{ij}) - C_{sync}\sin(2\phi_i(t)) \quad (A6.1)$$

Further, the energy function for this system can be described by:

$$E(\phi(t)) = -C_1 \sum_{i,j,\ j\neq i}^{N} J_{ij} \cos(\Delta\phi_{ij}) - \sum_{i=1}^{N} C_{sync}\cos(2\phi_i(t)) \quad (A6.2)$$

where $C_1$ is the coupling strength, and $C_{sync}$ modulates the strength of the second harmonic injection signal.

**Acknowledgment:**

This work was supported by NSF ASCENT grant (No. 2132918).